\documentclass{PoS}

\title{Discovery of TeV gamma-ray emission from the pulsar wind nebula 3C 58 by MAGIC}

\ShortTitle{PWN 3C 58 by MAGIC}

\author{\speaker{Rub\'en L\'opez-Coto}\\
        IFAE, Campus UAB, E-08193 Bellaterra, Spain\\
	E-mail: \email{rlopez@ifae.es}
        }

\author{Emiliano Carmona\\
        Centro de Investigaciones Energ\'eticas, Medioambientales y Tecnol\'ogicas, E-28040 Madrid, Spain\\
}  

\author{Wlodek Bednarek\\
        University of \L\'od\'z, PL-90236 Lodz, Poland\\
}        

\author{Oscar Blanch, Juan Cortina\\
        IFAE, Campus UAB, E-08193 Bellaterra, Spain\\
        }

\author{Emma de O\~na Wilhelmi\\
	Institute of Space Sciences, E-08193 Barcelona, Spain\\
}

\author{Diego Torres for the MAGIC Collaboration\\
        ICREA and Institute of Space Sciences, E-08193 Barcelona, Spain\\
}

\author{Jonatan Martin\\
	Institute of Space Sciences, E-08193 Barcelona, Spain\\
}

\author{Miguel \'A. P\'erez-Torres\\
      Inst. de Astrof\'isica de Andaluc\'ia (CSIC), E-18080 Granada, Spain \\
      also at Depto. de F\'isica Te\'orica, Facultad de Ciencias de la Universidad de Zaragoza, Spain\\
 now at Centro de F\'isica del Cosmos de Arag\'on, Teruel, Spain\\

}

\abstract{The pulsar wind nebula (PWN) 3C 58 has been proposed as a good candidate for detection at VHE (VHE; E>100 GeV) for many years. It is powered by one of the highest spin-down power pulsars known (5\% of Crab pulsar) and it has been compared to the Crab Nebula due to its morphology. This object was previously observed by imaging atmospheric Cherenkov telescopes (Whipple, VERITAS and MAGIC), and upper limit of 2.4\% Crab Unit (C.U.) at VHE. It was detected by Fermi-LAT with a spectrum extending beyond 100 GeV. We analyzed 81 hours of 3C 58 data taken with the MAGIC telescopes and we detected VHE gamma-ray emission with a significance of 5.7 sigma and an integral flux of 0.65\% C.U. above 1 TeV. We report the first significant detection of PWN 3C 58 at TeV energies. According to our results 3C 58 is the least luminous VHE gamma-ray PWN ever detected at VHE and the one with the lowest flux at VHE to date. We compare our results with the expectations of time-dependent models in which electrons up-scatter photon fields. The best representation favors a distance to the PWN of 2 kpc and Far Infrared (FIR) comparable to CMB photon fields. If we consider an unexpectedly high FIR density according to GALPROP, the data can also be reproduced by models assuming a 3.2 kpc distance. A low magnetic field, far from equipartition, is required to explain the VHE data. Hadronic contribution from the hosting supernova remnant (SNR) requires an unrealistic energy budget given the density of the medium, disfavoring cosmic ray acceleration in the SNR as origin of the VHE gamma-ray emission.

}

\FullConference{The 34th International Cosmic Ray Conference,\\
		30 July- 6 August, 2015\\
		The Hague, The Netherlands}

\begin{document}

\section{General description}
\label{intro}

The supernova remnant 3C 58 (SNR G130.7+3.1) has a flat radio spectrum and is brightest near the center, therefore it was classified as a pulsar wind nebula. It is centered on PSR J0205+6449, a pulsar discovered in 2002 with the {\it Chandra} X-ray observatory \cite{Discovery_X-rays}. It is widely assumed that 3C 58 is located at a distance of 3.2 kpc \cite{Distance_old}, but recent H I measurements suggest a distance of 2 kpc \cite{Kothes2013}. The age of the system is estimated to be $\sim$ 2.5 kyr \cite{Chevalier} from the PWN evolution and energetics, however this is a matter of debate. The pulsar has one of the highest spin-down powers known ($\dot E$ = 2.7$\times$10$^{37}$erg s$^{-1}$). The  PWN has a size of 9$^{\prime}$$\times$6$^{\prime}$ in radio, infrared (IR), and X-rays \cite{Size, Size_X-rays, Slane2004, IR_Mag_field}. Its luminosity is $L_{\textrm{ 0.5 -- 10 keV}}=2.4\times 10^{34}$ erg s$^{-1}$ in the X-ray band, which is more than 3 orders of magnitude lower than that of the Crab nebula \cite{X_Luminosity}. 3C 58 has been compared with the Crab because the jet-torus structure is similar \cite{Slane2004}. Because of these morphological similarities with the Crab nebula and its high spin-down power (5\% of Crab), 3C 58 has historically been considered one of the PWNe most likely to emit $\gamma$ rays.


The PWN 3C 58 was previously observed in the VHE $\gamma$-ray range by several IACTs. The Whipple telescope reported an integral flux upper limit of 1.31$\times$10$^{-11}$ cm$^{-2}$s$^{-1}$  $\sim$ 19 \% C.U. at an energy threshold of 500 GeV \cite{WhippleULs}, and VERITAS established upper limits at the level of 2.3 \% C.U. above an energy of 300 GeV \cite{3C58_Aliu_VERITAS}. MAGIC-I observed the source in 2005 and established integral upper limits above 110 GeV at the level of 7.7$\times$10$^{-12}$ cm$^{-2}$s$^{-1}$ ($\sim$4 \% C.U.)\cite{MAGICULs}. The improved sensitivity of the MAGIC telescopes with respect to previous observations and the {\it Fermi}-LAT results motivated us to perform deep VHE observations of the source.

\section{MAGIC observations and results} 
\label{3c58_results}
MAGIC observed 3C 58 in the period between 4 August 2013 to 5 January 2014 for 99 hours, and after quality cuts, 81 hours of the data were used for the analysis. The data were analyzed using the MARS analysis framework \cite{MARS}. The source was observed at zenith angles between 36$^\circ$ and 52$^\circ$. The data were taken in \emph{wobble-mode} \cite{Wobble} pointing at four different positions situated 0.4$^\circ$ away from the source to evaluate the background simultaneously with 3C 58 observations. 

\begin{figure}
\centering
\includegraphics[width=0.75\columnwidth]{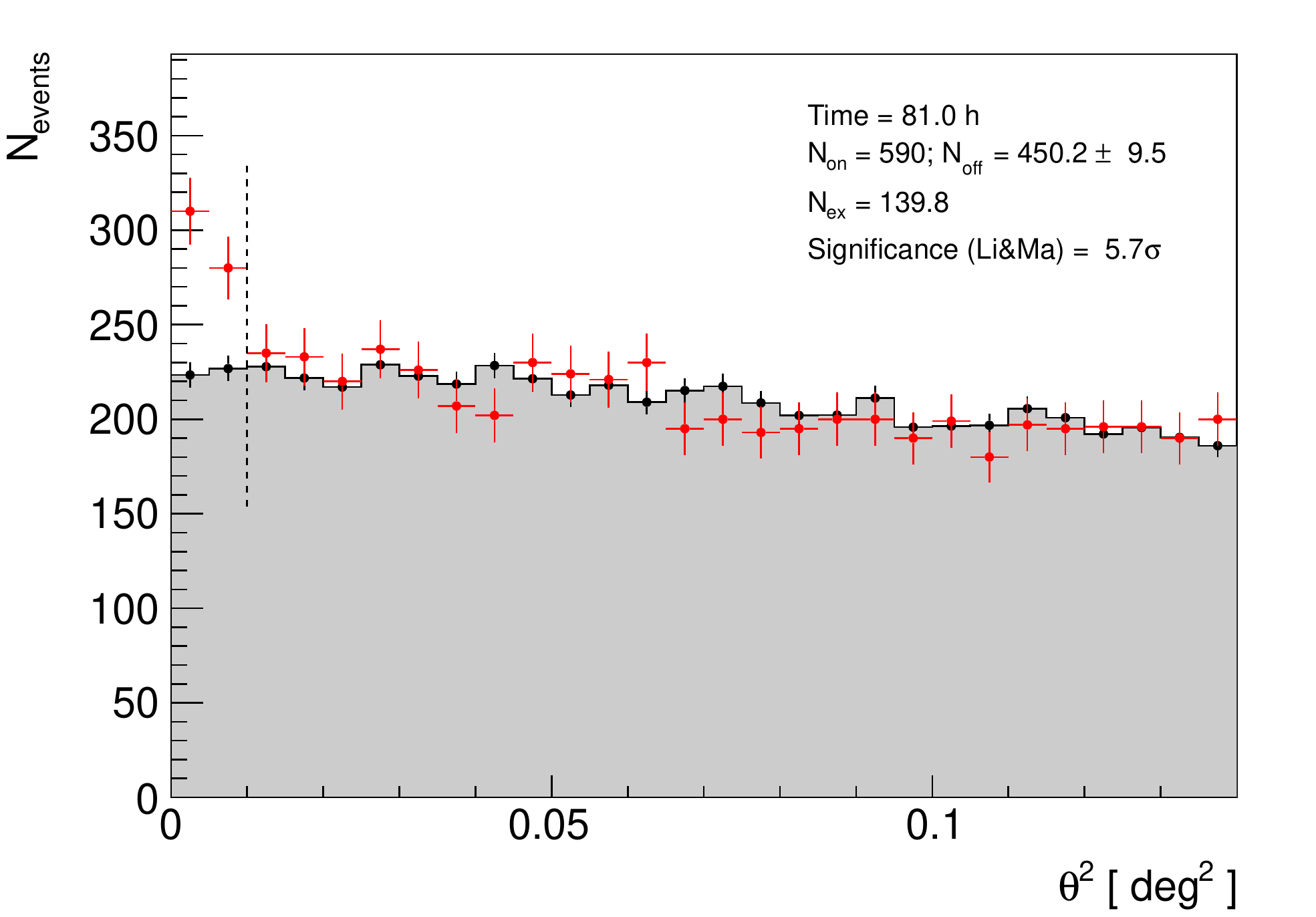}
\caption{Distribution of squared angular distance, $\theta^2$, between the reconstructed arrival directions of gamma-ray candidate events and the position of PSR J0205+6449 ({\it red points}). }
\label{Theta2}
\end{figure}

The applied cuts yield an energy threshold of 420 GeV. The significance of the signal, calculated with the LiMa formula, is $5.7\sigma$, which establishes 3C 58 as a $\gamma$-ray source. The $\theta^2$ distribution is shown in Figure \ref{Theta2}. As the five OFF positions were taken for each of the wobble positions, the OFF histograms were re-weighted depending on the time taken on each wobble position.

\begin{figure}[ht]
\vspace{-10pt}
\begin{center}
 \includegraphics[width=0.75\textwidth]{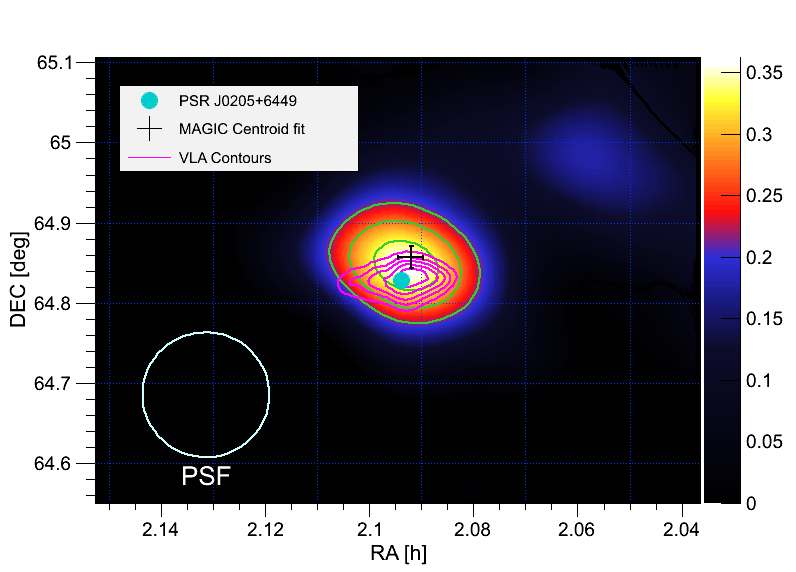}
\end{center}
\vspace{-20pt}
   \caption{Relative flux (excess/background) map for MAGIC observations. The cyan circle indicates the position of PSR J0205+6449 and the black cross shows the fitted centroid of the MAGIC image with its statistical uncertainty. In green we plot the contour levels for the TS starting at 4 and increasing in steps of 1. The magenta contours represent the VLA flux at 1.4 GHz \cite{VLA}, starting at 0.25 Jy and increasing in steps of 0.25 Jy.}
\vspace{-20pt}
  \label{Skymaps}
\end{figure}

We show in Figure \ref{Skymaps} the relative flux (excess/background) skymap, produced using the same cuts as for the $\theta^2$ calculation. The TS significance, which is the LiMa significance applied on a smoothed and modeled background estimate, is higher than 6 at the position of the pulsar PSR J0205+6449. The excess of the VHE skymap was fit with a Gaussian function. The best-fit position is RA(J2000)~=~2~h 05~m 31(09)$_\textrm{stat}$(11)$_\textrm{sys}$~s ; DEC~(J2000)~=~$64^\circ$ 51$^{\prime}$(1)$_\textrm{stat}$(1)$_\textrm{sys}$. This position is statistically deviant by $2\sigma$ from the position of the pulsar, but is compatible with it if systematic errors are taken into account. In the bottom left of the image we show the point spread function (PSF) of the smeared map at the corresponding energies, which is the result of the sum in quadrature of the instrumental angular resolution and the applied smearing (4.7$^{\prime}$ radius, at the analysis energy threshold). The extension of the signal is compatible with the instrument PSF. The VLA contours are coincident with the detected $\gamma$-ray excess.

\begin{figure}[ht]
 \centering
 \includegraphics[width=0.8\textwidth]{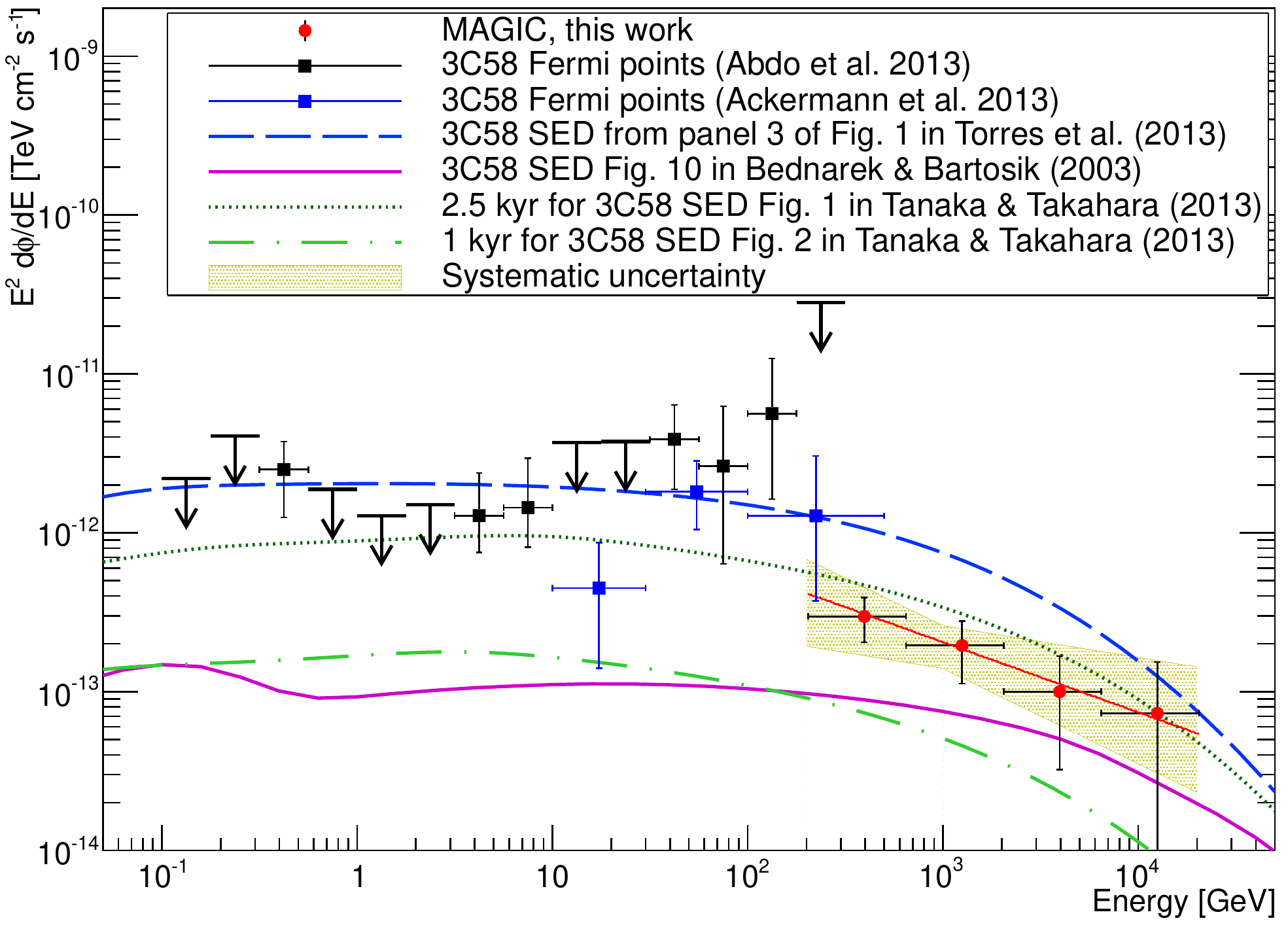}
\caption{3C 58 spectral energy distribution in the range between 0.1 GeV and 20 TeV. Red circles are the VHE points reported in this work. The best-fit function is drawn in red and the systematic uncertainty is represented by the yellow shaded area. Black squares and black arrows are taken from the {\it Fermi}-LAT second pulsar-catalog results \cite{SecondPulsarCatalog}. Blue squares are taken from the {\it Fermi} high-energy LAT catalog \cite{FermiAbove10}. The magenta line is the SED prediction for 3C 58 taken from Figure 10 of \cite{Wlodek2003}. The clear green dashed-dotted line is the SED predicted by \cite{Tanaka13}, assuming an age of 1 kyr, and the dark green dotted line is the prediction from the same paper, assuming an age of 2.5 kyr. The blue dashed line represents the SED predicted by \cite{3C58_Diego} assuming that the Galactic FIR background is high enough to reach a flux detectable by the MAGIC sensitivity in 50h.}
\vspace{-15pt}
  \label{SED}
 \end{figure}

Figure \ref{SED} shows the energy spectrum for the MAGIC data, together with published predictions for the gamma-ray emission from several authors, and two spectra obtained with three years of {\it Fermi}-LAT data, which were retrieved from the {\it Fermi}-LAT second pulsar-catalog (2PC) \cite{SecondPulsarCatalog}  and the {\it Fermi} high-energy LAT catalog (1FHL) \cite{FermiAbove10}. The 1FHL catalog used events from the {\it Pass 7 Clean class}, which provides a substantial reduction of residual cosmic-ray background above 10 GeV, at the expense of a slightly smaller collection area, compared with the  {\it Pass 7 Source class} that was adopted for  2PC \cite{Clean_class}. The two $\gamma$-ray spectra from 3C58 reported in the  2PC and  1FHL catalogs agree within statistical uncertainties. The differential energy spectrum of the source is well fit by a single power-law function d$\phi$/d$E$=$f_0$($E/$1 TeV)$^{-\Gamma}$ with  $f_{0} = (2.0 \pm 0.4_{\textrm{stat}} \pm 0.6_{\textrm{sys}})10^{-13} \textrm{cm}^{-2} \textrm{s}^{-1} \textrm{TeV}^{-1}$,  $\Gamma = 2.4 \pm 0.2_{\textrm{stat}}\pm 0.2_{\textrm{sys}}$ and $\chi^2$=0.04/2. The systematic errors were estimated from the MAGIC performance paper \cite{Performance2012} including the upgraded telescope performances. The integral flux above 1 TeV is $F_{E>1\textrm{ TeV}}=1.4\times10^{-13} \textrm{cm}^{-2} \textrm{s}^{-1}$. Taking into account a distance of 2 kpc, the luminosity of the source above 1 TeV is  $L_{\gamma, E>1 \textrm{ TeV}}=(3.0\pm1.1)\times$10$^{32}$$d^2_{2}$ erg s$^{-1}$, where $d_{2}$ is the distance normalized to 2 kpc.

\section{Discussion}



Several models have been proposed that predict the VHE $\gamma$-ray emission of PWN 3C 58. \cite{Wlodek2003} proposed a time-dependent model in which positrons gain energy in the process of resonant scattering by heavy nuclei. The VHE emission is produced by IC scattering of leptons off CMB, IR, and synchrotron photons and by the decay of pions due to the interaction of nuclei with the matter of the nebula. The age of 3C 58 is assumed to be 5 kyr, using a distance of 3.2 kpc and an expansion velocity of 1000 km s$^{-1}$. According to this model, the predicted integral flux above 400 GeV is $\sim$10$^{-13}$ cm$^{-2}$s$^{-1}$, while the integral flux above 420 GeV measured here is 5$\times$10$^{-13}$cm$^{-2}$s$^{-1}$. Calculations by \cite{Wlodek2005}, using the same model with an initial expansion velocity of 2000 km s$^{-1}$ and considering IC scattering only from the CMB, are consistent with the observed spectrum. However, the magnetic field derived in this case is $B\sim$14$\mu$G and it underestimates the radio emission of the nebula, although a more complex spectral shape might account for the radio nebula emission.

 \cite{Tanaka10} developed a time-dependent model of the spectral evolution of PWN including synchrotron emission, synchrotron self-Compton, and IC. They evolved the electron energy distribution using an advective differential equation. To calculate the observability of 3C 58 at TeV energies they assumed a distance of 2 kpc and two different ages: 2.5 kyr and 1 kyr \cite{Tanaka13}. For the 2.5 kyr age, they obtained a magnetic field $B\sim$17 $\mu$G, while for an age of 1 kyr, the magnetic field obtained is B=40 $\mu$G. The emission predicted by this model is closer to the {\it Fermi} result for an age of 2.5 kyr.

\cite{Jonatan} presented a different time-dependent leptonic diffusion-loss equation model without approximations, including synchrotron emission, synchrotron self-Compton, IC, and bremsstrahlung. They assumed a distance of 3.2 kpc and an age of 2.5 kyr to calculate the observability of 3C 58 at high energies \cite{3C58_Diego}. The predicted emission, without considering any additional photon source other than the CMB, is more than an order of magnitude lower than the flux reported here. It predicts VHE emission detectable by MAGIC in 50 hours for an FIR-dominated photon background with an energy density of 5 eV/cm$^3$. This would be more than one order of magnitude higher than the local IR density in the Galactic background radiation model used in GALPROP ($\sim$0.2 eV cm$^{-3}$) \cite{GALPROP}. The magnetic field derived from this model is 35 $\mu$G. To reproduce the observations, a large FIR background or a revised distance to the PWN of 2 kpc are required. In the first case, a nearby star or the SNR itself might provide the necessary FIR targets, although no detection of an enhancement has been found in the direction of the PWN. As we mentioned in Sec. \ref{intro}, a distance of 2 kpc has recently been proposed by \cite{Kothes2013} based on the recent H I measurements of the Canadian Galactic Plane Survey. At this distance, a lower photon density is required to fit the VHE data.



\section{Conclusions}

We have for the first time detected VHE $\gamma$ rays up to TeV energies from the PWN 3C 58. The measured luminosity and flux make 3C 58 into an exceptional object. It is the weakest VHE PWN detected to date, a fact that attests to the sensitivity of MAGIC. On the other hand, it is also the least luminous VHE PWN, far less luminous than the original expectations. Its ratio $L_{VHE}/\dot E \simeq 10^{-5}$ is the lowest measured, similar to Crab, which makes into a very inefficient $\gamma$-ray emitter. Only a closer distance of 2 kpc or a high local FIR photon density can qualitatively reproduce the multiwavelength data of this object in the published models. Since the high FIR density is unexpected, the closer distance with FIR photon density comparable with the averaged value in the Galaxy is favored. The models that fit the $\gamma$-ray data derived magnetic fields which are very far from equipartition.

\begin{acknowledgments}
\small
We would like to thank
the Instituto de Astrof\'{\i}sica de Canarias
for the excellent working conditions
at the Observatorio del Roque de los Muchachos in La Palma.
The financial support of the German BMBF and MPG,
the Italian INFN and INAF,
the Swiss National Fund SNF,
the ERDF under the Spanish MINECO (FPA2012-39502), and
the Japanese JSPS and MEXT
is gratefully acknowledged.
This work was also supported
by the Centro de Excelencia Severo Ochoa SEV-2012-0234, CPAN CSD2007-00042, and MultiDark CSD2009-00064 projects of the Spanish Consolider-Ingenio 2010 programme,
by grant 268740 of the Academy of Finland,
by the Croatian Science Foundation (HrZZ) Project 09/176 and the University of Rijeka Project 13.12.1.3.02,
by the DFG Collaborative Research Centers SFB823/C4 and SFB876/C3,
and by the Polish MNiSzW grant 745/N-HESS-MAGIC/2010/0.
\end{acknowledgments}


\bibliographystyle{JHEP}
\bibliography{references} 


\end{document}